\magnification\magstep1
\overfullrule = 0 pt
{\catcode`p =12 \catcode`t =12 \gdef\eeaa#1pt{#1}}	
\def\accentadj#1{\setbox0\hbox{$#1$}\kern	
		\expandafter\eeaa\the\fontdimen1\textfont1 \ht0 }
\def\overchar#1#2{\vbox{\ialign{$\hfil##\hfil$\crcr	%
	\accentadj{#2}#1\crcr\noalign{\kern.37ex\nointerlineskip}
	\displaystyle#2\crcr}}}
	
%
%

\def\SU{\mathcode`,="8000 \mathcode`!="8000\relax}	
{\catcode`\,=\active \catcode`\!=\active
\gdef,{\mkern-2mu{}_} \gdef!{\mkern-2mu{}^}}
\def\Ko{\oalign{$K$\cr	
	\noalign{\vskip0.3ex}	
	\hidewidth$\scriptstyle0$\hidewidth}{}}
\def\Ro{\oalign{$R$\cr
\noalign{\vskip0.3ex}
\hidewidth$\scriptstyle0$\hidewidth}{}}

\def\1{\oalign{$\hat{{\it L\,\,}}$\cr
\noalign{\vskip0.3ex}
\hidewidth$\xi_1$\hidewidth}{}}

\def\2{\oalign{$\hat{{\it L\,\,}}$\cr
\noalign{\vskip0.3ex}
\hidewidth$\xi_2$\hidewidth}{}}

 \def\Po{\oalign{$P$\cr
\noalign{\vskip0.3ex}
\hidewidth$\scriptstyle0$\hidewidth}{}}

\baselineskip 16pt

\def\R{\SU R^a,{bcd} }

\def\Po{$\SU \Po^a,{bcd}$ }

\def\Ko{$\SU \Ko^a,{bcd}$ }

\def\Ro{$\SU \Ro^a,{bcd}$ }

\def\R*{\check R}

\def\B*{\check B}

\def\Q*{\check Q}

\parindent=0pt

\def\PPh#1{\setbox0\hbox{$#1\rm I$}\mathord{\vcenter{\ialign{$#1\rm##$\cr
I\cr\noalign{\nointerlineskip \vskip-0.541\ht0}P\cr}}}}

\def\Ph{{\mathpalette\PPh{}}}
\def\Ed{{\mathord{\mkern5mu\mathaccent"7020{\mkern-5mu\partial}}}}

\chardef\Myxi="18

\parindent=0pt

\

{\bf INTEGRATION IN THE GHP FORMALISM III: FINDING ALL CONFORMALLY FLAT
RADIATION METRICS AS AN EXAMPLE OF AN `OPTIMAL SITUATION'.}

\ 

\ 

\

\centerline{S. Brian Edgar\footnote{$^1$}{ Department of Mathematics, 
 Link\"oping University, Link\"oping, Sweden S-581 83.} and Garry
Ludwig\footnote{$^2$}
{ Department of Mathematical Sciences, University of Alberta,
Edmonton, Alberta, Canada  T6G 2G1.} }

\

\

\

{\bf Abstract.}

\ 

Held has proposed an integration procedure within the GHP formalism built
around four real, functionally independent, zero-weighted scalars.
He suggests that such a procedure would be particularly simple for the `optimal
situation', when the formalism directly supplies the full quota of four scalars
of this type; a spacetime without any Killing vectors would be such a
situation. Wils has recently obtained a 
metric  which he claims is 
the only conformally flat, pure radiation
metric  which is not a plane wave; this metric has been shown by Koutras to admit
no Killing vectors, in general.  Therefore, as a  simple
illustration of the GHP integration procedure, we  obtain systematically the
complete class  of  conformally flat, pure radiation metrics.   Our result  
shows that the conformally flat, pure radiation metrics, which are not plane
waves, are a larger class than Wils has obtained.

\vfill\eject

{\bf 1. INTRODUCTION.}

About 20 years ago, soon after the introduction of the GHP formalism [1],
Held proposed ---
in principle --- a strikingly simple procedure  for integration within this
formalism:
manipulate the complete system of GHP equations with the aim of reducing them to  a
complete involutive set of tables for the action of the four GHP operators
$\Ph$, $\Ph'$,
$\Ed$, $\Ed'$ on six real GHP quantities, [2]. If such an  `optimal situation' 
can be achieved
Held emphasised that  the problem of
integrating the field
equations was essentially solved.

To avoid misunderstandings, we point out that the `complete system of GHP
equations' consists of the
GHP Ricci equations, the GHP Bianchi equations, and the GHP commmutator
equations, [3,4]:
while the `GHP quantities' consist of the GHP spin coefficients and the tetrad
components of the Riemann
tensor, together with any GHP operator-derivatives of these quantities
which arise in the
reduction procedure.  We now envisage the `optimal situation' more
precisely as follows:  a complete system of GHP equations reduces to a complete
involutive set of tables for the action of the four GHP operators on four real,
zero-weighted, functionally independent GHP quantities, and on one complex,
`weighted GHP
quantity', [4,5].  We will refer to the four real, zero-weighted,
functionally independent quantities  as `coordinate candidates' since they are the obvious, and
often the most convenient, choice for the coordinates when a coordinate
description of the
geometry and an explicit metric is required;  for a complex, `weighted GHP
quantity' we will require that neither spin nor boost weight of
this complex quantity is zero, i.e. $s\ne 0\ne t$, or equivalently
for GHP weights, $p\ne \pm q$ in the usual notations, [1].

Once we have obtained such a situation the problem is indeed solved, since
we can then use the tables to
 write down immediately  a coordinate description of the GHP operators $\Ph$,
$\Ph'$, $\Ed$,
$\Ed'$ (equivalently the tetrad vectors ${\bf l}$,${\bf n}$,${\bf m}$,${\bf
\bar m}$) in terms
of the coordinate candidates by applying the relationship for {\it
zero-weighted} GHP
quantities,
$$\nabla_i=l_i\Ph' +n_i\Ph - m_i\Ed' -\bar m_i\Ed
\eqno(1.1)$$
to each of the four coordinate candidates in turn.  The  complex
weighted quantity will, of
course, be cancelled out when the metric is constructed from the tetrad vectors by
$$g_{ij}=l_in_j + l_j n_i - m_i\bar m_j -m_j \bar m_i
\eqno(1.2)$$

Morever, if required, the explicit form of the tetrad vectors can be
simplified by using the
remaining freedom in the GHP formalism --- usually by gauging the complex
weighted quantity to
unity.

Surprisingly, until now, it has not been possible to construct an example
of such an optimal
situation, and so we have no explicit confirmation that Held's 
procedure works in practice.  Early
attempts, [2,5 - 9] to
integrate the GHP equations always failed to generate the full quota of
coordinate candidates; and so
in these cases it was thought necessary to introduce coordinates from
outside the GHP
formalism, and to complete the integration by reverting to the NP formalism
[10], and associated
explicit coordinate techniques [11,12]. 

We now understand better why those difficulties occurred; those early
investigations were carried out in
specialised spaces, which usually meant that they contained at least one
Killing vector. This, of
course, leads to at least one cyclic coordinate, and so the shortage of
explicit coordinate
candidates in the GHP approach is not surprising. (We emphasise that we are
{\it not} saying that the
presence of a Killing vector {\it always} leads to a shortage of coordinate
candidates; the
precise relationship between Killing vectors, tetrad vectors and coordinate
candidates needs careful
consideration, which we will present, in detail, elsewhere.)

However, recently, in such `less than optimal' situations where Killing
vectors are present, it
has been shown that it is possible to integrate the field equations {\it
completely within
the GHP formalism}, [4,13]; therefore, the coordinate- and gauge-free
spirit of Held's
original procedure is upheld, although in such situations extra quantities ---
additional to the
original GHP quantities --- have to be introduced in order to generate the
full quota of
coordinate candidates. (We should also mention some recent work by Kolassis 
[14],[15]
who has retained some of the spirit of Held's approach by using zero-weighted tetrad vectors,
although not zero-weighted coordinates, in general.)

So we have the apparent paradox: Held's approach promises to be simpler, at
least in principle, for
spaces without Killing vectors --- something contrary to our experience
with other approaches; on the
other hand, there does not exist an explicit example to illustrate this
supposedly simpler optimal
situation.

Recently, Wils [16] has obtained a comparatively  simple metric   and
Koutras [17] has pointed out that, in
general, this spacetime
contains no Killing vectors. In view of the discussion above, such a metric
is of interest in
the context of the GHP integration procedure.

Since Held's optimal situation suggests  the possibility of  a simple
procedure for investigating spacetimes without symmetries, we
decided to see whether this procedure does, in fact, work in practice for the
Wils metric.  So the original purpose of this paper was to  rederive Wils'
metric, and hence give an explicit demonstration of the GHP integration
procedure, in a comparatively
simple, but non-trivial, optimal situation. 

However, it turns out that --- from the same starting point of conformally flat,
pure radiation metrics --- we obtain a more general metric than Wils; in
fact  our expression includes the metrics of Wils [16] and Koutras and McIntosh
[18] respectively, as special cases. This illustrates precisely one advantage of
our GHP integration method compared to the familiar NP [10 - 12] coordinate-based
method. In the latter method, a careful account of coordinate and tetrad freedom
is required, which is gradually used up, often in  a  considerable number of
different steps, involving long tortuous calculations; not surprisingly,
sometimes  something   is overlooked. For this particular case, as Wils has
pointed out, Kramer et al. [19] mistakenly concluded that the only conformally
flat pure radiation metric were the plane waves, found by McLenaghan et al.
[20]; Wils [16], in turn, mistakenly concluded that his metric   --- from the
Kundt class [21] but not representing plane waves ---  completed the class
of    conformally flat  pure radiation metrics, which are not plane waves.
However, because of the simplicity of our calculations, and their susceptibility
to easy confirmation,  we are able to  state 
 unambiguously that our  more general metric   completes  this
 class of    conformally flat,  pure radiation metrics,
which are not plane waves. Its determination, in the familiar way,  from the
Kundt form of the metric, has been reported in [22]. Independent confirmation 
that the metric presented here is indeed more general than the Wils metric [16]
has been given by  Skea [23], who has used the invariant classification of the
CLASSI program [24].  The detailed
classification by Skea of this metric, and of the Wils metric [16],  is available
at the on-line exact solutions database in Brazil and in Canada, [23]. 

\

In Section 2 we  give a simple step-by-step illustration of our method,
for conformally flat, pure radiation metrics which are not plane waves 
--- in the generic case. 
In this situation we are able to choose all four of
our coordinate candidates directly from the GHP quantities --- precisely as
Held had envisaged;  but this of course involves the additional constraint that
none of these four quantities is constant. In  Section 3, by a slight
modification, we  find it is  easy to obtain our complete
class of these conformally flat pure radiation metrics --- the generic case 
together with  the excluded special
cases --- in
 one explicit expression. In this situation we are able to choose three of
our coordinate candidates directly from the GHP quantities, and the remaining
coordinate candidate indirectly.
 
 In Section 4 we present the
complete metric in the more familiar Kundt form, [19,21]; we also give an
alternative version, in a form which retains some coordinate freedom. 
In Section 5 we discuss further the principles and advantages of
our method. 

As a preliminary step, at the beginning of the calculation, it is advantageous to
fix (almost) completely the freedom of the two real null vectors ${\bf l},{\bf
n}$; the details of this choice are given in the Appendix.
 
For those familiar with the NP coordinate integration procedures, [10 - 12] it
may help to emphasise the fundamental differences in the approach illustrated for
the GHP formalism in this paper.  In order to obtain the metric for a
particular class of spacetimes, the NP approach begins by making a tentative
choice of a metric form, in a preferred coordinate system  and tetrad  frame
(suggested by the geometry of the class), and determines what coordinate and
tetrad freedom exists to retain this tentative metric form. Next, the three sets
of NP equations --- the Bianchi equations, the Ricci equations and the metric
equations  ---  are written down in these coordinates, and they are integrated
step-by-step; to facilitate the integration, the coordinate and tetrad freedom
is gradually used up, making the metric more precise, but still within the
tentative form. Furthermore, the tetrad and coordinate freedom are linked
together --- a change in one, usually necessitates a change in the other.

In the approach in this paper, there is no tentative preliminary choice of metric
nor of coordinate system; instead the three sets of GHP equations --- the Bianchi
equations, the Ricci equations and the commutator equations  --- are immediately
simplified, within the  operator notation --- a process which is
equivalent to integrating the field equations. Once this process is completed 
--- in the form of six
tables --- the coordinates are chosen  in terms of zero-weighted GHP quantities;
in the optimal situation, these coordinates are chosen directly and uniquely,
with no coordinate freedom. The metric only occurs at the very last step, and
can be written down from the tables, in terms of these coordinates. {\it We
emphasise that in our approach  the metric form has been dictated by the nature
of the calculations themselves, and that only zero-weighted GHP quantities are
chosen as coordinates.}  Furthermore, the remaining GHP
tetrad freedom --- built into the single weighted complex quantity of our
analysis --- is completely decoupled from our coordinate candidates, since they
are zero-weighted; changing the GHP gauge does not affect the  coordinate
candidates. 

In this particular application we fixed (almost) completely the two real null
vectors ${\bf l},{\bf n}$ at the beginning of the calculation; but this is not
essential, and the calculation can be carried out with some, or all, of the
freedom of these two null rotations; but in such cases the final 
metric may not be in its simplest or neatest form.

\vfill\eject

\beginsection 2. THE INTEGRATION PROCEDURE: THE GENERIC CASE.

{\bf 2.1.  Preliminary simplifications.}

We consider the conformally flat spacetimes with energy momentum tensor given by
$$T_{ij} = \Phi^2 l_i l_j
\eqno(2.1)$$
where $\Phi$ is a scalar function, and $l_i$ is a null vector.  If we
identify this null vector with
its counterpart in the  usual  null tetrad ${\bf l}$,${\bf n}$,${\bf
m}$,${\bf \bar m}$ we obtain,
in usual GHP notation,
$$\Psi_0=\Psi_1=\Psi_2=\Psi_3=\Psi_4=0\eqno(2.2a)$$
$$\Phi_{00}=\Phi_{11}=\Phi_{01}=\Phi_{02}=\Phi_{12}=\Lambda=0, \qquad\qquad
\Phi_{22}=\Phi^2\eqno(2.2b)$$
When these values are substituted into the GHP Bianchi equations, it
follows immediately that
$$\sigma = \rho = \kappa=0 \eqno(2.3)$$
\medskip
{\it At this stage it is necessary to subdivide our problem into two cases: $(i)\  
\tau=0,\  (ii) \  \tau \ne 0$. In this paper we shall restrict our explicit attention
to the second of these cases.} The first of these cases corresponds to a subset
of the plane wave spacetimes [19,21].
\medskip
We still have the freedom to choose the tetrad vector
${\bf n}$ up to a null rotation about ${\bf l}$, and exploiting this freedom
 we can choose (see Appendix),
$$\tau' = \sigma' = \rho' = 0 \eqno(2.4a)$$ 
$$\Phi^2-\tau \kappa' - \bar \tau \bar \kappa' =
0, \eqno(2.4b)$$ $$\Ph'(\tau / \bar \tau ) = 0, \eqno(2.4c)$$
Since the Bianchi equations for $\Phi_{22}$ are identically satisfied under
the substitution
(2.4b), the only remaining equations from the Ricci and Bianchi equations GHP
equations are,
$$\eqalign{\Ph \tau & = 0\qquad\qquad\qquad\qquad \ \ \Ph \kappa' = 0\cr 
\Ed \tau & = \tau^2 \qquad\qquad\qquad\qquad\   \Ed \kappa' = - \bar \tau \bar \kappa'
\cr \Ed'\tau & = \tau \bar \tau \qquad\qquad\qquad\qquad \Ed' \kappa' =\bar \tau \kappa'
}\eqno(2.5)$$

The commutators become
$$
\eqalign{[\Ph,\Ph'] & =\bar \tau \Ed + \tau \Ed'\cr
[\Ed,\Ed'] & =0 \cr
[\Ph,\Ed] & =0 =[\Ph,\Ed']  \cr
[\Ph',\Ed] & =-\bar \kappa' \Ph - \tau \Ph'\cr
[\Ph',\Ed'] & = -\kappa' \Ph - \bar \tau \Ph'} \eqno(2.6)$$
It is very important to note that in order to extract {\it all} the
information from these commutator
equations, they must be applied explicitly to {\it all} four coordinate
candidates, and to a weighted
complex quantity, [3].

\medskip

\

{\bf 2.2. Finding four coordinate candidates, and extracting all the information
from the complete system.}

The spin coefficients and Riemann tensor components therefore supply four
real quantities (from
complex $\tau,\kappa'$) which can easily be rearranged into two real
zero-weighted and one complex
weighted quantity.  The simplest zero-weighted quantities would appear
   to be $\tau\bar\tau$ and
$\tau\kappa'/\bar \tau \bar \kappa'$, while an obvious weighted quantity --- especially in view of the gauge
choice (2.4c) --- is
$(\tau/\bar
\tau)$; these quantities satisfy respectively,
$$\eqalign{\Ed (\tau \bar \tau) & =  2 \tau^2 \bar \tau \cr
\Ed (\tau \kappa' / \bar \kappa'\bar \tau) & =  - \tau \bigl(1+ \tau \kappa'/
\bar \tau
\bar \kappa'\bigr)\cr 
\Ed(\tau /\bar \tau) &  =  0}\eqno(2.7)$$
and from these quantities it is straightforward to construct a set of tables.
However, to keep the presentation of subsequent calculations to a mininimum,
it will be convenient to begin 
instead with the two real combinations of these two  zero-weighted quantities,
$$A={1\over\sqrt{2\tau\bar\tau}}
\eqno(2.8a)$$
$$B=
{i(\tau\kappa'-\bar\tau\bar\kappa')\over(\tau\kappa'+\bar\tau\bar\kappa')\sqrt{2\tau\bar
\tau}} = iA{-1+(\tau \kappa' / \bar \kappa'\bar \tau)\over 
1 + (\tau \kappa' / \bar \kappa'\bar \tau)}
 \eqno(2.8b)$$
and with the complex weighted quantity $(PQ) $ (with GHP weights $(0,1)$),
given by, 
$$P=\sqrt{\tau\over2\bar\tau}
 \qquad\qquad \hbox{with}\qquad  P\bar P ={1\over 2}\eqno(2.9a)$$

 $$Q={\sqrt{\tau\kappa'+\bar\tau\bar\kappa'}\over\root 4 \of
{2\tau\bar\tau}}
\eqno(2.9b)$$
 (Note
that from (2.4b) the term $\tau\kappa'+\bar\tau\bar\kappa'$ is positive.)

These particular
choices of $A,B, P,Q $ have been made because they enable us to replace
(2.7) with the very simple equations
$$\eqalign{
\Ed A & =  -P  \qquad\quad\qquad\qquad\qquad \Ed B =-iP\cr
}\eqno(2.10)$$
$$\eqalign{ \Ed P & = 0  \qquad\qquad\qquad\qquad\qquad \Ed Q =0}\eqno(2.11)$$

(We have already assumed from the beginning of this section that $\tau\ne0$, and
clearly in the gauge (2.4b), $\kappa' \ne 0$; therefore $A,B,P,Q$ will always be
defined, and  differ from  zero.)

Complete tables for $A,B,P,Q$ can now be presented,
$$\eqalign{\Ph A & =  0  \qquad\qquad\qquad\qquad\qquad \Ph B =0\cr
\Ed A & =  -P  \qquad\quad\qquad\qquad\qquad \Ed B =-iP\cr
\Ed' A & =  -\bar P  \quad\qquad\qquad\qquad\qquad \Ed' B =i\bar P\cr
\Ph' A & =  QC/A  \qquad\qquad\qquad\qquad \Ph' B =QE/A\cr
\  & \  \cr
\Ph P & = 0
\qquad\qquad\qquad\qquad\qquad \Ph Q =0\cr
\Ed P & = 0  \qquad\qquad\qquad\qquad\qquad \Ed Q =0\cr
\Ed' P & = 0  \qquad\qquad\qquad\qquad\qquad \Ed' Q =0\cr
\Ph' P & = 0  \qquad\qquad\qquad\qquad\qquad \Ph' Q =Q^2G/A}\eqno(2.12)$$
by introducing the three real zero-weighted quantities
$C,E,G$ respectively --- as yet undetermined.  (We have introduced  the factor
${Q\over A}$ in the above definitions simply for convenience in later
calculations.)

Since neither $A$ nor $B$ can be constant, and also since
$$
\nabla A \ne \lambda \nabla B\eqno(2.13)
$$
for any scalar $\lambda$, clearly $A$ and $B$ are functionally independent,
and can be adopted as
coordinate candidates. Therefore, we next  have to apply the commutators (2.6) to
$A,B$, and also to the
weighted $P,Q$. This results in the non-trivial equations respectively,
$$\eqalign{\Ph C & = -1/Q  \qquad\qquad\qquad \Ph E
=0\qquad\qquad\qquad\qquad \Ph G
=0\cr
\Ed C & = 0  \qquad\qquad\qquad\qquad\Ed E =0\qquad\qquad\qquad\qquad \Ed G
=0\cr
\Ed' C & = 0  \qquad\qquad\qquad\qquad \Ed' E =0\qquad\qquad\qquad\qquad
\Ed' G =0
}\eqno(2.14)
$$

\medskip

At this stage, we still require two more coordinate candidates, in addition to
$A,B
$, to make up our full quota.
Since  $C$ cannot be constant, and also since
$$
\nabla C \ne \lambda \nabla B+\mu \nabla B\eqno(2.15)
$$
for any scalars $\lambda,\mu$, clearly  $C$ is functionally independent of
$A$ and $B$, and can be
adopted as the third coordinate candidate. So we therefore obtain a table
for $C$,
$$\eqalign{\Ph C & = -1/Q \cr
\Ed C & = 0  \cr
\Ed' C & = 0  \cr
\Ph' C & = QJ/A}\eqno(2.16)$$
 which is completed with the  real zero-weighted quantity
$J$  --- as yet undetermined.

When we apply the commutators (2.6) to the third coordinate candidate $C$ we
obtain, $$\eqalign{ \Ph J & = G/Q\cr
 \Ed J & = P(A+iB)\cr
 \Ed' J & =\bar P(A-iB)
}\eqno(2.17)$$
Rearranging  we define the real zero-weighted quantity $S$ by
$$\eqalign{S=J+CG+(A^2+B^2)/2}\eqno(2.18)$$
so that
$$\eqalign{\Ph S & = 0 \cr
\Ed S & = 0  \cr
\Ed' S & = 0  }\eqno(2.19)$$

The obvious choice for our fourth coordinate candidate is $E $; but of course
that is only possible if $E $ is not constant.
\medskip
{\it In the remainder of this section we shall consider only the generic case
where $E$ is not a constant.}
\medskip
So we therefore complete a table for $E$,
$$\eqalign{\Ph E & = 0 \cr
\Ed E & = 0  \cr
\Ed' E & = 0  \cr
\Ph'E & =QH/A}\eqno(2.20)$$
where the real, zero-weighted quantity
$H$ is as yet undetermined. 

A check on the
determinant formed from the four tables for $A,B,C,E$ respectively shows
that all four quantities are
functionally independent, and so we can adopt $A,B,C,E$ as our four
coordinate candidates.

The only information in the GHP field equations still unused is now
obtained by applying the
commutators to the last  coordinate candidate $E$, obtaining 
$$\eqalign{\Ph H & = 0  \cr
\Ed H & = 0   \cr
\Ed' H & = 0  
}\eqno(2.21)$$

So we have  extracted {\it all} the information from the GHP Ricci, Bianchi and
commutator equations.

\

{\bf 2.3. The six tables.}

We now have the following six tables,
$$\eqalign{\Ph A & =  0  \qquad\qquad\qquad\qquad\qquad\qquad\qquad\qquad \Ph B
=0\cr
\Ed A & =  -P  \qquad\quad\qquad\qquad\qquad\qquad\qquad\qquad \Ed B =-iP\cr
\Ed' A & =  -\bar P  \quad\qquad\qquad\qquad\qquad\qquad\qquad\qquad \Ed' B
=i\bar P\cr
\Ph' A & =  QC/A  \qquad\qquad\qquad\qquad\qquad\qquad\qquad \Ph' B =QE/A\cr
\  & \ \cr
\Ph C & =-1/Q\qquad\qquad\qquad\qquad\qquad\qquad\qquad\Ph E  = 0\cr
\Ed C & = 0  \qquad\qquad\qquad\qquad\qquad\qquad\qquad \qquad  \Ed E=0\cr
\Ed' C & = 0  \qquad\qquad\qquad\qquad\qquad \qquad\qquad \qquad \Ed' E =0\cr
\Ph' C & = Q(S-CG-{1\over 2}A^2-{1\over
2}B^2)/A \qquad\qquad\Ph' E  =  QH/A\cr
\ & \ \cr
\Ph P & = 0  \qquad\qquad\qquad\qquad\qquad\qquad\qquad\qquad \Ph Q =0\cr
\Ed P & = 0  \qquad\qquad\qquad\qquad\qquad\qquad\qquad\qquad  \Ed Q =0\cr
\Ed' P & = 0  \qquad\qquad\qquad\qquad\qquad\qquad\qquad\qquad  \Ed' Q =0\cr
\Ph' P & = 0  \qquad\qquad\qquad\qquad\qquad\qquad\qquad\qquad \Ph' Q
=Q^2G/A}\eqno(2.22)$$ where the real zero-weighted quantities $G,H,S$ satisfy
$$\eqalign{
\Ph G & = 0 = \Ed G\cr  \Ph H & = 0 = \Ed H\cr\Ph S & = 0
= \Ed S\cr }\eqno(2.23)$$

Strictly speaking the six tables (2.22) are not involutive since they have to be
supplemented
by (2.23). However, it follows from (2.23) that $G,H,S$ are functions only of
the one coordinate candidate, $E$ and so by stipulating that
$G,H,S$  be functions {\it only} of 
$E$, we no longer need to write out (2.23)
explicitly, and the tables are essentially involutive; hence the problem is
essentially solved.

\
{\bf 2.4. Using coordinate candidates as coordinates.}

We now make an obvious choice of the coordinate candidates as the coordinates,
$e,c,a,b$,
$$  e=E,\qquad c=C, \qquad a=A, \qquad b=B
\eqno(2.24)$$
Using (1.1) as follows
$$l^e =l^i\nabla_i(e) = \Ph (E) \qquad\qquad \hbox{etc.}
$$
we can write down the tetrad vectors immediately in the $e,c,a,b$
coordinates from the respective tables as
$$\eqalign{l^i & = (0,{-1\over Q},0,0)\cr
n^i & = {Q\over a}\Bigl(H,(S-Gc-{1\over 2}a^2-{1\over
2}b^2),c,e\Bigr)\cr
m^i & = P(0,0,-1,-i)\cr
\bar m^i & = \bar P(0,0,- 1,i)\cr
}\eqno(2.25)$$
 and the metric  is given by,
$$g^{ij}=\pmatrix{0 &{-H/ a} & 0&0\cr
{-H/ a} & (-2S+2Gc+a^2+b^2)/a & -{c/ a} & {-e/ a}\cr
0&-{c/ a} & -1&0\cr
0&{-e/ a}&0&-1}
\eqno(2.26)$$
where $G,H,S$ are arbitrary functions  of the coordinate $e$; clearly $H$
cannot be zero.

This has completed our integration procedure, and we have obtained the metric 
for all spacetimes satisfying (2.1,2), which are subject to the restrictions $\tau\ne
0$, and --- with respect to the gauge chosen here ---  $E$ not a
constant.

\vfill\eject

\beginsection 3. THE INTEGRATION PROCEDURE: THE COMPLETE METRIC.

{\bf 3.1. Preliminaries.}

In the previous section 2.2 we assumed that $E$ was not a constant, so that we
were able to choose it as our fourth coordinate candidate. Next, we should
look at the excluded case where $E$ is a constant. In such a situation, clearly
$H$ is zero, but we still have the possibility of choosing $G$ or $S$ as our
fourth coordinate. Once we make such a choice then we could continue in a similar
manner as in the last section, building our tables, and hence the tetrad, around
our four coordinate candidates. However, if {\it all} of the functions $E,G,S$
are constants, then it will {\it not} be possible to find the fourth coordinate
candidate  directly; we emphasise that in such circumstances no additional
independent quantities can be generated by any direct manipulations of the
tables and the commutators. In such a situation we still need a fourth coordinate
candidate in order to extract all the information  from the commutators. 
Clearly some indirect way is required to obtain this information. We shall now
show that such an indirect approach to the fourth coordinate can in fact be used
in general, so that we can obtain the complete metric as one expression.

\

{\bf 3.2. Finding a fourth coordinate candidate indirectly, and extracting all
the information from the complete system.}

The results in Section 2 up to (2.19) apply, and therefore when we write out
our tables explicitly we obtain (2.22) --- except that the table for the
quantity $E$ is missing.

Clearly we do not have our full quota of {\it four} coordinate candidates, but
we do not wish to use any of the remaining quantities in the five tables,
since it would involve the additional condition of that quantity being
non-constant.  However, we know that we have not yet extracted all the
information from the commutators (2.6), since they have only been applied to
{\it three} zero-weighted coordinate  candidates. So
we examine the commutators, 
$$ \eqalign{[\Ph,\Ph'] &
=(\bar P/A) \Ed + ( P/A) \Ed'\cr [\Ed,\Ed'] & =0 \cr [\Ph,\Ed] & =0 =[\Ph,\Ed'] 
\cr [\Ph',\Ed] & =( P Q^2(A+iB)/A) \Ph - ( P/A) \Ph'\cr
[\Ph',\Ed'] & = (\bar P Q^2(A-iB)/A) \Ph - (\bar P/A) \Ph'} \eqno(3.1)$$
  to determine whether they suggest the existence of a
 fourth zero-weighted quantity, functionally independent of the first three
coordinate candidates, whose table is consistent with the commutators.
In fact, we get a strong hint from the previous section, and consider the
possibility of the existence of a real zero-weighted quantity $T $, which
satisfies the table  
$$\eqalign{\Ph T & =  0\cr
 \Ed T & =  0\cr
\Ed' T & =  0\cr
\Ph' T & =  Q/A
}\eqno(3.2)$$
It is
straightforward to confirm that such a choice is consistent with the
commutators and the other five tables in (2.22).
Furthermore, a check on the determinant formed from the four tables for $A,B,C,T$
respectively, shows that all four quantities are functionally independent.

\ 

{\bf 3.3 The six tables.}

Therefore, in the set of tables (2.22), we can replace  the table for $E$ with
the  table (3.2) for $T$, and the real zero-weighted quantities $E,G,S$
satisfy
$$\eqalign{
\Ph E & = 0 = \Ed E\cr  \Ph G & = 0 = \Ed G\cr\Ph S & = 0
= \Ed S\cr }\eqno(3.3)$$

\

{\bf 3.4 Using coordinate candidates as coordinates.}

We now make the obvious choice of the coordinate candidates as the coordinates,
$$ t=T,\qquad  c=C, \qquad a=A, \qquad b=B
\eqno(3.4)$$
where the only coordinate freedom is for $t$ up to an additive constant.
We can write down the tetrad vectors immediately in the $t,c,a,b$
coordinates from the respective tables as
$$\eqalign{l^i & = (0,{-1\over Q},0,0)\cr
n^i & = {Q\over a}\Bigl(1,(S-G c-{1\over 2}a^2-{1\over
2}b^2),c,E \Bigr)\cr
m^i & = P(0,0,-1,-i)\cr
\bar m^i & = \bar P(0,0,- 1,i)\cr
}\eqno(3.5)$$
and therefore the metric  is given by,
$$g^{ij}=\pmatrix{0 &{-1/ a} & 0&0\cr
{-1/ a} & (-2S+2G c+a^2+b^2)/a & -{c/ a} & -{E/ a}\cr
0&-{c/ a} & -1&0\cr
0&-{E / c}&0&-1}
\eqno(3.6)$$
where $E,G,S$ are arbitrary functions of the coordinate $t$. This form now
includes the possibility of $E$ (or G, or S) being constant.

\vfill\eject

\beginsection 4. ALTERNATIVE FORMS FOR THE COMPLETE METRIC.

Although the expression (3.6) is in a concise form,  we can make a
coordinate transformation,
$$u = U(t) \qquad\quad v=-ca/V(t), \quad\qquad x=-a,\quad\qquad
y=b+W(t)/2 \eqno(4.1)
$$ 
which will take  the metric into the familiar Kundt form [19,21], in coordinates
$u,v,x,y$,

$$g^{ij}= \pmatrix{0 &{\dot U\over V } & 0&0\cr
{\dot U\over V } & Z  & -{2v \over x} & (2E+\dot
W)\over 2V\cr  0& -{2v\over x} &-1&0\cr  0& (2E+\dot
W)\over 2V&0&-1} \eqno(4.2)$$
where 
$$Z = -{3v^2\over x^2} -{2v\over V^2}(\dot V+VG) -{x\over V^2
}\Bigl(-2 S+ (W^2/4) -Wy + x^2+y^2 \Bigr)
$$

and \ $\dot{}$ \  denotes differentiation with respect to $t$.

Choosing
$$W(e)=-2\int{E} de, \qquad V(e)=\exp(-\int {G } de),
\qquad U(e)=\int{V}
de,
\quad
\eqno(4.3)$$
we find the metric in the $u,v,x,y$
coordinates becomes,
$$g^{ij}=\pmatrix{0&{1} & 0 & 0\cr {1} & -2 f(u)x\Bigl( x^2+y^2
+g(u) y  +h(u)\Bigr) -{3v^2/ x^2}& -{2v/ x} &
0\cr0& -{2v/ x} & -1&0\cr 0&0&0&-1}
\eqno(4.4)$$
where
 $f(\ne0),g,h  
$ are arbitrary functions  of
the coordinate
$u$. 
(There are of course other possibile ways to exploit this coordinate transformation.)

\

We emphasise that in our method we do not
always have to fix our coordinates completely as we did in Section 3 and above.
Instead, we could have introduced the fourth coordinate  with some freedom via
the table for $\tilde T$, given by $$\eqalign{\Ph \tilde T & =  0\cr
 \Ed \tilde T & =  0\cr
\Ed' \tilde T & =  0\cr
\Ph' \tilde T & =  Q \tilde H /A
}\eqno(4.5)
$$
where $\tilde H$ is an abitrary quantity, as yet undetermined. By this approach 
we could have obtained the complete metric in $\tilde t(=\tilde T),c,a,b$
coordinates, $$g^{ij}=\pmatrix{0 &-{\tilde H(\tilde t)/ a} &0&0 \cr
-{\tilde H(\tilde t)/ a} & \Bigl(-2\tilde S(\tilde t)+2\tilde G(\tilde
t)c+a^2+b^2\Bigr)/a &-{c/a} & -{E(\tilde t)/ a}\cr 0&-{c/a} & -1&0\cr
0&-{E(\tilde t)/ a}&0&-1}
\eqno(4.6)$$
where $E,\tilde G,\tilde H,\tilde S$ are now arbitrary functions  of 
the coordinate $\tilde t$. There remains the freedom 
$\tilde t\rightarrow
F(\tilde t)$ where $F$ is an arbitrary function of $\tilde t$ which can be 
used to choose: (i) $\tilde t=E(\tilde
t)$ (if
$E$ is not constant, as in Section 2); or (ii) $\tilde t$ as some other of the 
(non-constant) quantities
$\tilde G,\tilde S,\tilde H$; or (iii)
$\tilde H=1$ (as in Section 3); or (iv) other convenient possibilities. 

\

For completeness we present the covariant form of (4.4) in $u,v,x,y$
coordinates,
$$g_{ij}=\pmatrix{  2f(u)x\Bigl(x^2+y^2 +g(u) y 
+h(u)\Bigr) -{v^2/ x^2} &{  1} &{-2v/
x} & 0\cr 
 1 & 0 & 0 & 0\cr {-2v/ x} &0& -1&0\cr
0&0&0&-1} 
\eqno(4.7)$$ 
This metric --- subject to a difference in sign convention ---  agrees with the
form given in [22], and Wils' special case [16] is obtained when $g=0=h$.    

\

\vfill\eject

\beginsection 5. DISCUSSION.

The metric given in (3.6) --- equivalently (4.4)  --- represents the
complete class of all conformally flat, pure radiation metrics, which are not
plane waves.

However, the primary purpose of this paper was to illustrate our method, and
demonstrate how it differs from the more familiar NP tetrad approach. One
important aspect of this approach is the crucial role of the commutators as
field equations in their own right, and the need to extract {\it all} the
information from them in a systematic manner; in fact, in this example very
little information came directly from the Ricci and Bianchi equations, and the
major part via the commutators.  In addition, we wished to emphasise the fact
that in this method, ideally, the coordinates are fixed directly and completely
in terms of zero-weighted GHP quantities --- probably the most significant
difference from the NP approach; this was illustrated explicitly in Section 2 in
the choice of all four coordinate candidates.  In Section 3, the situation was a
little different because we choose not to 
 equate  our fourth coordinate  candidate directly to a GHP quantity; 
rather,  the fourth coordinate   candidate $T$ had to be introduced as a
`potential' (essentially an integral) of GHP quantities. But, also in this case,
the coordinate candidate was chosen uniquely ---  without any
coordinate freedom, except for an additive constant. 

To emphasise that in our method we are not always bound  to fix our coordinates
completely, we showed in Section 4 how we can permit a measure
of coordinate freedom; although with the dangers and complications inherent in
introducing such coordinate freedom, it is often better to fix the coordinates
where possible.

In this paper we have retained the term `coordinate candidates' for our four
real functionally independent scalars, although, in fact, we do eventually choose
them  as our coordinates.  We use this term because we prefer to distinguish
between the role of the coordinate candidates in extracting all the information
from the commutators, and their possible additional and optional role as
coordinates in the final explicit statement of the metric. Especially in less
than optimal situations, it may not be convenient to choose the coordinate
candidates as the eventual coordinates.

In those special circumstances when the formalism does not directly yield
four coordinate candidates --- for instance when E,G,H are all constants in (3.6)
--- the spacetime admits a Killing vector, since  in this case, the fourth
coordinate candidate is cyclic. Of course, this does not mean that we can
conclude that  Killing vectors are absent in other cases, for the general metric.
The explicit links between Killing vectors, tetrad vectors and the existence of
coordinate candidates in the GHP formalism will be considered elsewhere.

\ 

Although this particular example, which we have chosen to illustrate our GHP
integration method, underlines the simplicity and conciseness of the method
compared to the NP coordinate approach, we emphasise that these are not our only
reasons for developing this method. As Held has pointed out, and demonstrated in
[6,7], such a method has the potential to extract additional information when
other methods have been brought to a stop. In addition, the procedure followed
has much in common with aspects of the Karlhede classification of spacetimes
[25]; in fact, once a spacetime has been obtained by the method in this paper,
its Karlhede classification --- by a GHP approach similar to that  introduced by
Collins et al. [26] --- is a comparatively simple undertaking.

\

{\bf ACKNOWLEDGEMENTS.}

One author (BE) would like to thank the Mathematics Department of the
University of Alberta for its hospitality while part of this work was
being carried out, and for travel support from the Swedish Natural
Science Research Council. He would also like to thank Fredrik Andersson for
discussions. The other author (GL) is grateful for the continuing financial
support by the Natural Sciences and Engineering Research Council of Canada. We
also thank Alan Held for his critical reading of the first version of this
paper, and for  continuing discussions.

\parindent=0pt

\vfill\eject

\beginsection APPENDIX

Assuming (2.2,3), under the null rotation
$$\eqalign{l^i & \rightarrow  l^i \cr
m^i & \rightarrow  m^i +Z l^i\cr
\bar m^i & \rightarrow  \bar m^i +\bar Z l^i\cr
n^i & \rightarrow  n^i +Z m^i+\bar Z \bar m^i + Z\bar Z l^i}
\eqno(A1)$$
the zero valued spin coefficients 
$\kappa, \rho, \sigma$ do not change, while
$$\eqalign{
\tau & \rightarrow  \tau \cr 
\tau' & \rightarrow  \tau'- \Ph Z \cr
\rho' & \rightarrow  \rho'- \Ed Z -\bar Z(\tau'- \Ph Z) \cr
\sigma' & \rightarrow  \sigma'- \Ed' Z - Z(\tau'- \Ph Z)   \cr
\kappa' & \rightarrow  \kappa' - \Ph' Z - Z(\rho' - \Ed Z)-\bar Z (\sigma'- \Ed' Z)
 -\bar Z
 Z (\tau'- \Ph Z) - Z^2 
\tau \cr
\Phi_{22} & \rightarrow  \Phi_{22} }
\eqno(A2)$$
We could therefore choose $Z$ such that $\tau',\rho',\sigma',\kappa'$ are all 
zero, 
{ providing the choices} 
$$\eqalign{
\tau'- \Ph Z & = 0\cr
  \rho'- \Ed Z  & = 0\cr
   \sigma'- \Ed' Z & = 0  \cr
  \kappa' - \Ph' Z  - Z^2 
\tau  & = 0
 }\eqno(A3)$$
{ are consistent with the commutator equations for $Z$}; but this is {\it not}
the case.

However, a careful examination of the calculations leads us to note that we can
choose 
$$\eqalign{
\tau'- \Ph Z & = 0\cr
  \rho'- \Ed Z  & = 0\cr
   \sigma'- \Ed' Z & = 0   \cr
 \Phi_{22}-\tau \kappa' -\bar\tau \bar \kappa' + \tau \Ph' Z + \bar\tau \Ph' \bar Z + Z^2 \tau^2 + \bar Z^2 \bar\tau^2  & = 0
 }\eqno(A4)$$
{ since these choices are consistent with the relevant commutators for $Z$.}

(As an illustration, we consider the commutator 
$$
[\Ph,\Ed]Z =  \bar \tau' \Ph Z  
$$
and the substitution of the first two equations of (A4) results in
$$\Ph \rho' - \Ed \tau' +\bar \tau' \tau =0
$$
which is one of the Ricci identities. For all other commutators, we also get one of the Ricci identities.)

There is clearly some freedom left in our choice for $Z$, and noting that,  
$$
\Ph' \rightarrow \Ph' + Z\Ed + \bar Z \Ed' + \bar Z Z \Ph- p Z \tau - q \bar Z \bar \tau
\eqno(A5)$$
we see that
$$
\Ph'(\tau/\bar \tau) \rightarrow  \Ph' (\tau/\bar \tau)
-2(Z\tau-\bar Z\bar \tau)\tau/\bar \tau
\eqno(A6)$$

If we choose $Z$ such that
$$
  \Ph' (\tau/\bar \tau)
-2(Z\tau-\bar Z\bar \tau)\tau/\bar \tau = 0
\eqno(A7)$$
we find that this choice is consistent with the previous  choices (A4), for $Z$.

We have therefore fixed the behaviour of all four operators on the type $(
-2,0)$ quantity $Z$, 
as well as fixing part of $Z$ itself by $(A7)$; the result being that we can
choose
$$\eqalign{
 \tau' = \rho' = \sigma' & = 0\cr
 \Phi_{22}-\tau \kappa' -\bar\tau \bar \kappa'   & = 0\cr
 \Ph' (\tau/\bar \tau) & = 0 }
\eqno(A8)$$ (In an appropriate gauge,  we have essentially fixed $Z$ up to a
real  constant.)

\  

{\bf REFERENCES.}

1. Geroch R., Held A., and Penrose R. (1973). {\it J. Math. Phys.,} {\bf 14,}
874. 

2. Held A. (1974). {\it Commun. Math. Phys.,} {\bf 37,} 311.

3. Edgar S. B. (1980). {\it Gen Rel. Grav.,} {\bf 12,} 347.

4. Edgar S. B. (1992). {\it Gen Rel. Grav.,} {\bf 24,} 1267.

5. Held A. (1985). In {\it Galaxies, Axisymmetric Systems and
Relativity} (ed. M.A.H. MaCallum), Cambridge University Press. p.208.

6. Held A. (1975). {\it Commun. Math. Phys.,} {\bf 44,} 211.

7. Held A. (1976). {\it Gen Rel. Grav.,} {\bf 7,} 177.

8. Held A. (1976). {\it J. Math. Phys.,} {\bf 17,} 39.

9. Stewart J.M. and Walker M. (1974). {\it Proc.
Roy.  Soc. A} {\bf 341,} 49.

10. Newman E.T. and Penrose, R. (l962). {\it J. Math. Phys.,} {\bf 3,} 566.

11. Newman E.T. and Unti, T. (1962). {\it J. Math. Phys.,} {\bf 3}, 891.

12. Newman E.T. and Unti, T. (1963). {\it J. Math. Phys.,} {\bf 4}, 1467.

13. Edgar S. B. and  Ludwig G. (1996).    `Integration in the GHP formalism II:
An operator approach with applications to twisting Type N spaces.' {\it 
 Gen. Rel. Grav.}, in press.

14. Kolassis Ch. (1996). {\it Gen Rel. Grav.} {\bf 28,} 787.

15. Kolassis Ch. (1996). {\it Gen Rel. Grav.} {\bf 28,} 805.

16. Wils P. (1989). {\it Class. Quantum Grav.} {\bf 6,} 1243.

17. Koutras A. (1992). {\it Class. Quantum Grav.} {\bf 9,} L143.

18. Koutras A. and McIntosh C.  (1996). {\it Class. Quantum Grav.} {\bf 13,}
L47

19. Kramer D., Stephani H., MacCallum M. and Herlt E. (1980). {\it Exact
Solutions of Einstein's Equations.} (Cambridge: Cambridge University Press).

20. McLenaghan R. G., Tariq  N and Tupper B. O. J. (1975). {\it J. Math. Phys.,}
{\bf 16}, 829.

21. Kundt W. (1962). {\it Proc. Roy. Soc. Lond. } {\bf A 270,} 328.

22. Edgar S. B. and  Ludwig G. (1996). `All conformally flat pure radiation
metrics.' {\it Preprint:  GR-GC/9612059 }

23. Skea J. (1996). {\it private communication}, and at

http://edradour.symbcomp.uerj.br  \  and \  
 http://www.astro.queensu.ca/\~{}jimsk.

24. \AA man J.E. (1982).  {\it Manual for  CLASSI: Classification program for
geometries in General Relativity.} Preprint,  University of Stockholm.

25. Karlhede A. (1980). {\it Gen Rel. Grav.} {\bf 12,} 693.

26. Collins J. M., d'Inverno R. A. and Vickers J. A. (1990). {\it Class.
Quantum Grav.} {\bf 7,} 2005.

\end